%% file: arxiv.tex
\documentclass[
reprint,
amsmath,amssymb,
aps,
pra
]{revtex4-2}

\usepackage{graphicx}
\usepackage{dcolumn}
\usepackage{bm}
\usepackage{physics}
\usepackage{braket}
\usepackage{placeins}
\usepackage[colorlinks=true]{hyperref}
\usepackage{upgreek}
\usepackage{multirow}
\usepackage{dsfont}
\usepackage{lipsum}
\usepackage{booktabs}
\begin{document}

\let\oldaddcontentsline\addcontentsline% Store \addcontentsline
\renewcommand{\addcontentsline}[3]{}% Make \addcontentsline a no-op

\input{main_header.tex}

\date{\today}

\maketitle

\input{main.tex}
\bibliography{library}

\FloatBarrier
\clearpage
\onecolumngrid
\begin{center}
{\Large \textbf{Supplementary information}}
\end{center}
\makeatletter
   \renewcommand\l@section{\@dottedtocline{2}{1.5em}{2em}}
   \renewcommand\l@subsection{\@dottedtocline{2}{3.5em}{2em}}
   \renewcommand\l@subsubsection{\@dottedtocline{2}{5.5em}{2em}}
\makeatother
\let\addcontentsline\oldaddcontentsline% Restore \addcontentsline

\renewcommand{\thesection}{\arabic{section}}

\twocolumngrid

\let\oldaddcontentsline\addcontentsline% Store \addcontentsline
\renewcommand{\addcontentsline}[3]{}% Make \addcontentsline a no-op
\let\addcontentsline\oldaddcontentsline% Restore \addcontentsline
\renewcommand{\theequation}{S\arabic{equation}}
\renewcommand{\thefigure}{S\arabic{figure}}
\renewcommand{\thetable}{S\arabic{table}}
\renewcommand{\thesection}{S\arabic{section}}
\setcounter{figure}{0}
\setcounter{equation}{0}
\setcounter{section}{0}

\newcolumntype{C}[1]{>{\centering\arraybackslash}p{#1}}
\newcolumntype{L}[1]{>{\raggedright\arraybackslash}p{#1}}

% \tableofcontents
\FloatBarrier

\input{SI.tex}

\end{document}

%% file: main_header.tex
\title{Polarisation-insensitive state preparation for trapped-ion hyperfine qubits}

\begin{abstract}
    Quantum state preparation for trapped-ion qubits often relies on high-quality circularly-polarised light, which may be difficult to achieve with chip-based integrated optics technology. 
    We propose and implement a hybrid optical/microwave scheme for intermediate-field hyperfine qubits which instead relies on frequency selectivity. 
    Experimentally, we achieve $99.94\%$ fidelity for linearly-polarised ($\sigma^+$/$\sigma^-$) light, using $^{43}$Ca$^+$ at 28.8 mT.
    We find that the fidelity remains above $99.8\%$ for a mixture of all polarisations ($\sigma^+$/$\sigma^-$/$\pi$).
    We calculate that the method is capable of $99.99\%$ fidelity in $^{43}$Ca$^+$, and even higher fidelities in heavier ions such as $^\text{137}$Ba$^\text{+}$.
\end{abstract}

\author{A.\,D.\,Leu}
\thanks{These authors contributed equally}

\author{M.\,C.\,Smith}
\thanks{These authors contributed equally}

\author{M.\,F.\,Gely}
\thanks{These authors contributed equally}

\author{D.\,M.\,Lucas}

\affiliation{Clarendon Laboratory, Department of Physics, University of Oxford, Parks Road, Oxford OX1 3PU, U.K.}

%% file: main.tex
%!TEX root = arxiv.tex

Trapped ions are one of the foremost experimental platforms for quantum computation~\cite{IonTrapScaling}, demonstrating leading error figures for single-qubit gates~\cite{harty2014,leu2023} and entangling gates~\cite{srinivas2021,Clark2021}.
Ions also make excellent quantum memories, especially isotopes with nuclear spin, which feature ``clock'' qubit transitions with coherence times that can exceed an hour~\cite{Sepiol2019,Wang2021}.
Using ``clock'' qubits comes however at the cost of a dense hyperfine level structure, complicating motional cooling and qubit state preparation and measurement (SPAM)~\cite{quantinuumStatePrep,Ballance2016,Edmunds2021,Allcock2016}.
This is particularly true for isotopes with nuclear spin $I>1/2$, for example the commonly used $^\text{9}$Be$^\text{+}$~\cite{Gaebler2016}, $^\text{25}$Mg$^\text{+}$~\cite{srinivas2021,Tan2015}, $^\text{43}$Ca$^\text{+}$~\cite{Allcock2016,harty2014} or $^\text{137}$Ba$^\text{+}$ ions~\cite{quantinuumStatePrep}.
In this work we focus on state preparation in such isotopes.
Optical pumping schemes~\cite{OpticalPumping} are required to prepare these isotopes into specific hyperfine states.
For $I>1/2$ isotopes, optical pumping typically relies on polarisation selection rules to render one state dark to laser driving.
The success of such an approach hinges on high polarisation purity of the optical pumping beam.
However, this requirement constrains the beam delivery setup, and can complicate the use of on-chip integrated optics~\cite{Mehta2020}.
Whilst integrated delivery of high-purity circularly~\cite{Massai2022} and linearly~\cite{Mehta2017} polarised light has been demonstrated, it will introduce an additional degree of complexity when scaling quantum processors.
\begin{figure}[!ht]
    \centering
    \includegraphics[width=0.45\textwidth]{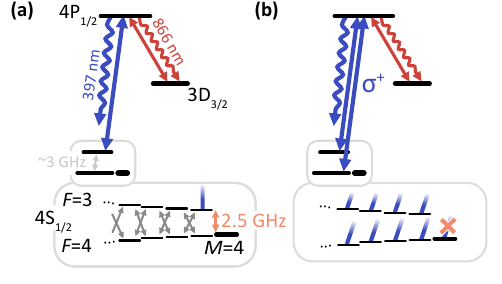}
    \caption{
    Both panels show the energy levels of a $^{43}$Ca$^{+}$ ion at 28.8 mT and the transitions involved in the preparation of the ``stretch'' state $4$S$_{1/2}$ $\ket{F=4,M=+4}$.
    \textbf{(a)} \textbf{{Fr}equency-selective state preparation (FSSP)}.
    A 397 nm laser is used to pump population out of the $\ket{F = 3,M= +3}$ state, coupled through microwave driving (grey arrows) to all states in the 4S$_{1/2}$ manifold except $\ket{F=4,M=+4}$.
    The hyperfine splitting provides a 2.5 GHz frequency detuning to laser transitions involving $\ket{F=4,M=+4}$, rendering the latter state almost dark to the optical pumping process.
    \textbf{(b)} \textbf{Polarisation-selective state preparation (PSSP)}.
    Here the 397 nm light targets both hyperfine manifolds of 4S$_{1/2}$ through the use of an electro-optical modulator.
    Using purely $\sigma^+$-polarised light leaves the $\ket{F=4,M=+4}$ state dark to this optical pumping process (indicated by the orange cross).
    In both schemes, population decaying to the 3D$_{3/2}$ manifold is repumped using an 866 laser (red).
    }
    \label{fig:main_fig1}
\end{figure}

Recently, an alternative approach to state-preparation of $I>1/2$ isotopes has been demonstrated which does not solely rely on high-purity polarisation~\cite{quantinuumStatePrep}.
In Ref.~\cite{quantinuumStatePrep}, optical pumping enables the preparation of a mixture of states within one hyperfine manifold, where the other manifold remains dark due to the frequency detuning arising from the hyperfine splitting.
Coherent microwave or quadrupole laser pulses are then used to transfer all but one state back to the bright hyperfine manifold for another round of optical pumping, and then the cycle is repeated.
Avoiding leakage out of the prepared state imposes a minimum duration of the coherent pulses, and polarisation-dependent optical pumping is thus used to reduce the total state preparation duration.
To eliminate the requirements on polarisation altogether, without compromising on the state-preparation duration, one could introduce a larger Zeeman shift between the hyperfine levels to isolate the target state in frequency space.
This could be achieved by using an ``intermediate-field'' clock qubit, such as the 14.6 or 28.8~mT clock transition in Ca$^+$, or the 21.3~mT transition in Mg$^+$.
Regardless of state-preparation considerations, the advantages of intermediate-field qubits have been used to demonstrate some of the highest performance logical gates, in particular using microwave (MW) driven gates~\cite{harty2014,harty2016,srinivas2021,Tan2015,Gaebler2016}.
However, we find that the scheme of Ref.~\cite{quantinuumStatePrep} is not effective at intermediate-field, resulting in percent-level state-preparation errors (see Sec.~\ref{sec:comparison}), prompting the alternative scheme demonstrated in this work.
\begin{figure}[!ht]
    \centering
    \includegraphics[width=0.45\textwidth]{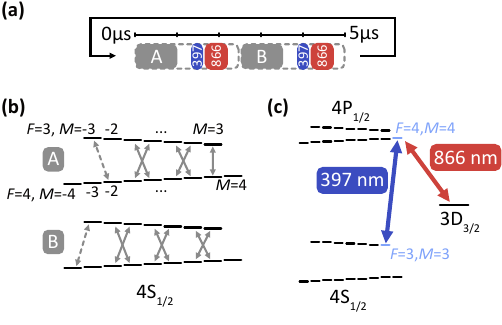}
    \caption{
    \textbf{FSSP pulse sequence.}
    \textbf{(a)} Sequence of MW (grey), 397 nm laser (blue), and 866 nm laser (red) pulses used for FSSP.
    \textbf{(b)} During the sequence, we alternate between driving different combinations of microwave transitions (A and B), within the hyperfine structure of the 4S$_{1/2}$ manifold.
    Weaker MW transitions (dashed grey) are driven for longer durations, as shown by the grey dashed box in (a).
    \textbf{(c)} The 397 nm laser pulse is centred on the 4S$_{1/2}$ $\ket{F=3,M=+3} \rightarrow$ 4P$_{1/2}$ $\ket{F=4,M=+4}$ transition and the 866 nm laser pulse drives power-broadened transitions to repump population that decays to the 3D$_{3/2}$ manifold.
    }
    \label{fig:main_fig2}
\end{figure}

In this Letter, we present a frequency selective state preparation (FSSP) method relying on the large Zeeman shifts of ``intermediate-field'' $I>1/2$ ions and their hyperfine splitting rather than laser polarisation purity.
We demonstrate state preparation of a qubit state in 838 $\upmu$s with a SPAM error of $1.2 (1) \times 10^{-3}$, and discuss the expected errors when FSSP is applied to other ion species.
We compare this technique to an implementation of a polarisation selective state preparation (PSSP) scheme.
%
%state prep time, add 88us for transfer pulses (initial pumping 500 us+20 us pumping + 5 cycles x (20 us pumping + 2*2us pi pulse (+ 100 us delay between pi pulses needed)) (1228 us)
%
Using the PSSP scheme, we demonstrate an improvement on previous implementations~\cite{harty2014}, reaching a SPAM error of $4.7(4) \times 10^{-4}$ with a state-preparation duration of $1.2$ ms.
The two techniques are shown schematically in Fig.~\ref{fig:main_fig1}.
Experiments are carried out on a segmented-electrode surface trap with an on-chip MW resonator, described in Ref.~\cite{weber2022cryogenic}.
Single $^{43}$Ca$^{+}$ ions are trapped at a 40 $\upmu$m height at room temperature.
Our qubit is defined by the hyperfine states $\ket{F=3,M=+1}=\ket{0}$  and $\ket{F=4,M=+1}=\ket{1}$ in the 4S$_{1/2}$ ground state manifold.
A quantisation field of 28.8 mT makes this transition insensitive to magnetic fields (to first order), yielding a so-called ``clock'' qubit.
State-preparation of a qubit state is achieved in two steps.
The focus of this work is the first step, which consists of preparing the $\ket{F=4,M=+4}$ state of the 4S$_{1/2}$ ground state manifold, which we refer to as the ``stretch'' state $\ket{s}$.
The second step is to use MW pulses to transfer the population from the stretch state to the $\ket{0}$ qubit state.
To readout, MW pulses map the qubit states $\ket{0}$ and $\ket{1}$ to $\ket{s}$ and $\ket{F=3,M=+1}$ respectively, after which optical pulses transfer or ``shelve'' the $\ket{s}$ state to states in the 3D$_{5/2}$ manifold, enabling state-dependent fluorescence~\cite{myerson2008}.
\begin{table}[b]
    \centering
    \setlength\doublerulesep{0.2cm}
        \begin{tabular}{|c||c|c|}
         \hline
         Error ($\times 10^{-5}$) &$\ket{1}$ (bright)&$\ket{0}$ (dark)\\
         \hline\hline
         Leakage & $4(4) $ & $5(2)$ \\
         \hline
         Decoherence & $4.02(14)$ & $7.1(2)$ \\
         \hline
         Detuning & $0.14(12)$ & $0.3(2)$ \\
         \hline
         Amplitude miscalibration & $6.2(9)$ & $6.5(9)$ \\
         \specialrule{1.5pt}{0pt}{0pt}
         \textbf{Total transfer pulse error} & $14(4)$ & $19(2)$ \\
         \hline
         \hline
         Off-resonant shelving\textit{*} & $25$ & - \\
         \hline
         Thresholding & $3.4(10)$ &  $3.4(10)$ \\
         \hline
         Shelving failure\textit{*} & - &  $4.0$ \\
         \hline
         Deshelving\textit{*} & - & $7.6$ \\
         \specialrule{1.5pt}{0pt}{0pt}
         \textbf{Total optical readout error} & $28.4(10)$ & $15.0(10)$ \\
         \hline
         \hline
         \textbf{Total expected error} & $42(6)$ & $34(2)$ \\
         \hline
         \textbf{Measured error} & $49(5)$ & $45(5)$ \\
         \hline
        \end{tabular}
        \\
    	\raggedright
        \textit{*Simulation}
\caption{
    %\textbf{Error budget.}
%
This error budget shows an estimate of the errors arising from the MW transfer pulses and the optical readout.
}
\label{table:error_budget}
\end{table}

In Table~\ref{table:error_budget}, we summarise our characterisation of the transfer pulse and readout errors, which is necessary to isolate the error in preparing $\ket{s}$ from the total SPAM error.
The transfer pulse errors are dominated by (i) leakage to other states occurring during MW oscillator frequency updates, (ii) decoherence during the pulses induced by magnetic field fluctuations, and (iii) accuracy of the pulse amplitude calibration.
The readout is limited by (i) off-resonant shelving of the bright state, (ii) spontaneous decay from the shelf states and (iii) dark counts due to undesired scattering of the 397 nm beam~\cite{Wineland80}.
The characterisation of these errors is detailed in Supplementary Secs.~\ref{sec:transfer_pulses},\ref{sec:optical_readout}.
We first discuss the use of a PSSP scheme to prepare the $\ket{s}$ state, illustrated in Fig.~\ref{fig:main_fig1}(b).
This relies on purely circularly-polarised 397 nm light to excite the 4S$_{1/2}$ to 4P$_{1/2}$ transition, leaving the $\ket{s}$ state unaffected as it possesses no $\Delta M = +1$ transition to 4P$_{1/2}$.
As a result, the $\ket{s}$ state becomes the steady-state of this optical pumping process.
An 866 nm laser is used to repump any population that decays to the 3D$_{3/2}$ states.
To bridge the hyperfine splitting within the 4S$_{1/2}$ manifold, we use an electro-optical modulator (EOM), where the EOM-generated sideband drives population out of the 4S$_{1/2}$ $F=4$ manifold.
As in Ref.~\cite{harty2014}, the effect of slight impurity in the 397 nm laser polarisation is corrected for by driving population out of the 4S$_{1/2}$ $F=3$ manifold only (by switching off the EOM) in combination with the MW transfer pulses $\ket{F=4,M=+3}\rightarrow\ket{F=3,M=+3}$ and $\ket{F=4,M=+2}\rightarrow\ket{F=3,M=+2}$.
Using PSSP, we measure a SPAM error of $1.4(2) \times 10^{-4}$ for the $\ket{s}$ state (without transfer pulses) which is consistent with the optical readout error for the dark state (see Table~\ref{table:error_budget}), indicating that the PSSP of $\ket{s}$ introduces only a negligible error.
When introducing transfer pulses, we measure a SPAM error (averaged over both qubit states) of $4.7(4) \times 10^{-4}$, which is in reasonable agreement with the error budget in Table~\ref{table:error_budget}.
As discussed in the introductory paragraphs, restrictions on optical access to the ion or limited control over polarisation (for example when using integrated optics) can prohibit the use of PSSP, prompting the FSSP scheme summarised in Fig.~\ref{fig:main_fig1}(a).
With this scheme, we extend the MW pulse step implemented at the end of the PSSP~\cite{harty2014} to remove the reliance on polarisation purity altogether.
We first exploit the hyperfine splitting ($\sim$3 GHz) which greatly exceeds the 397 nm transition linewidth ($\sim$23 MHz), enabling optical pumping out of the $F$=3 manifold of 4S$_{1/2}$ with limited effect on the $F$=4 states.
Secondly, we rely on the large Zeeman shift between hyperfine states ($\sim$100 MHz) to implement microwave transfer pulses between the $F$=3 and $F$=4 manifold of 4S$_{1/2}$ at a $\sim$1 MHz Rabi rate without affecting the state to be prepared ($\ket{s}$).
\begin{figure}[!ht]
    \centering
    \includegraphics[width=0.45\textwidth]{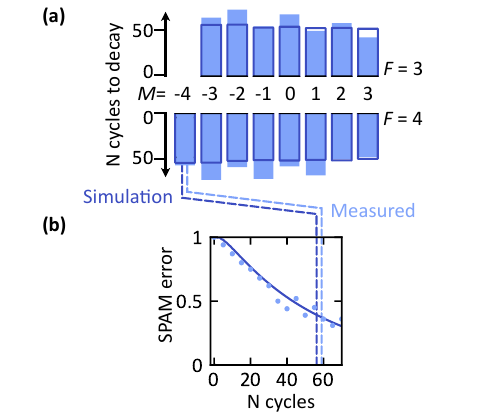}
    \caption{
    \textbf{FSSP: Comparison to numerical predictions.}
    \textbf{(a)} Number of cycles required to reach a SPAM error of $1/e$, for all possible initial states.
    Experimental results (light blue bars) are compared to expected values (dark blue) calculated through a rate equation model, and show predictable behaviour.
    From left to right, and top to bottom, bars correspond to states prepared with higher $M$ and $F$ number respectively.
    \textbf{(b)} Dataset used to produce one bar in panel (a).
    In this example, the state $\ket{F=4,M=-4}$ is prepared initially.
    A varying number of FSSP cycles is applied, before the state $\ket{F=4,M=4}$ is read out, revealing the SPAM error (light blue data points).
    The measured results are in excellent agreement with the simulated curve (dark blue).
    }
    \label{fig:main_fig3}
\end{figure}
\begin{figure}[!ht]
    \centering
    \includegraphics[width=0.45\textwidth]{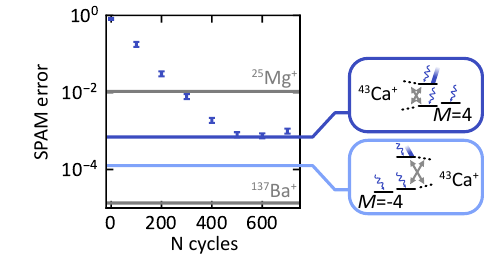}
    \caption{
    \textbf{FSSP performance in $^{43}$Ca$^{+}$ and other ion species.}
    The state-preparation and measurement (SPAM) error is measured for a varying number of FSSP cycles (dark blue dots), and is found to be in good agreement with the simulated fundamental limit (dark blue line).
    %
    % For each data point displayed in the graph, data is collected until the uncertainty in its value falls below 10$\%$ of its nominal value.
    %
    We compare this data to the expected state-preparation error if $\ket{F=4,M=-4}$ was prepared rather than $\ket{F=4,M=4}$, revealing the gains coming from a more favourable Zeeman shift.
    The grey lines show the expected state-preparation errors for different ion species, showing how the error decreases roughly quadratically with an increased hyperfine splitting, making FSSP particularly well suited to heavier ions such as Ba.
    }
\label{fig:main_fig4}
\end{figure}

In Fig.~\ref{fig:main_fig2}, the pulse sequence and driving scheme for the FSSP method is illustrated in more detail.
The scheme is designed to facilitate comparison to a numerical model, and, to this end, we avoid coupling three or more states together with different driving terms.
Avoiding such $\Lambda$-type systems allows us to straightforwardly simulate this process using rate equations, see Sec.~\ref{sec:Simulation}.
This is achieved by applying the 397 nm and 866 nm laser pulses sequentially and by using two groups of microwave pulses, also applied sequentially.
To simplify the microwave driving, we use the fact that each pair of transitions $M \rightarrow M+1$ and $M +1\rightarrow M$ are separated by only $\sim 160 $ kHz, and therefore we can drive these transitions simultaneously.
Due to the on-chip resonator, centred around our qubit frequency (3.123 GHz), certain transition frequencies (light grey in Fig.~\ref{fig:main_fig2} (b)) are strongly suppressed, so we drive these transitions for longer pulse times.
We first verify that we can drive population out of every state in the ground state manifold.
To do so, a given state is prepared using PSSP and transfer pulses, the FSSP scheme is applied for a varying number of cycles, and then the population in the stretch state is read out through shelving and fluorescence detection.
As the data in Fig.~\ref{fig:main_fig3} shows, the scheme is successful regardless of the starting state, and in good agreement with numerical predictions.
We then assess the steady-state state-preparation error.
As shown in Fig.~\ref{fig:main_fig4}, after 600 cycles (3 ms) of FSSP, state-preparation of $\ket{s}$ is achieved with an error of $6.1 (11) \times 10^{-4}$.
Here we have subtracted the optical readout error presented in Table~\ref{table:error_budget} from the total $7.5 (11) \times 10^{-4}$ error measured.
Again, this error is in good agreement with the numerical prediction of $6.3 \times 10^{-4}$ state-preparation error.
The FSSP error is limited by off-resonant excitation of $\ket{s}\rightarrow$4P$_{1/2}$ transitions.
This error scales with $1/\Delta ^{2}$, where $\Delta$ is the frequency difference between the 397 nm laser and the frequency of such transitions.
Taking into account both the hyperfine splitting and Zeeman shift, we have $\Delta>2.5$ GHz for all transitions.
A first route to lowering the state preparation error would be to prepare the $\ket{F=4,M=-4}$ state instead, where the sign of the Zeeman shift works in our favour, increasing the detuning to $\Delta>3.9$ GHz.
We predict an error for the preparation of $\ket{F=4,M=-4}$ of $1.3 \times 10^{-4}$.
However, in our experimental setup, the on-chip resonator suppresses the amplitude  of the transfer pulses necessary to transfer $\ket{F=4,M=-4}$ to a qubit state.
As a consequence, these transfer pulses would be significantly slower than those we typically use, resulting in a SPAM error limited by decoherence affecting these pulses.
A larger hyperfine splitting is therefore advantageous, making this technique particularly well suited to heavier ions.
We extend our simulation to other isotopes at ``intermediate-field'', \textit{i.e.} at static fields where the first clock transition which is not $M=0\leftrightarrow M=0$ appears in the ground-state.
As expected, we find that the error worsens for a lighter ion (we predict $1.1 \times 10^{-2}$ for preparing the $\ket{F=3,M=-3}$ state in $^{25}$Mg$^{+}$) and improves for a heavier ion ($1.4 \times 10^{-5}$ error for preparing the $\ket{F=2,M=2}$ state in $^{137}$Ba$^{+}$).
These predictions are shown in Fig.~\ref{fig:main_fig4} alongside our experimental results.
Finally, another route to lowering the error is to purify the polarisation of the 397 nm light, as, through selection rules, this limits the probability of off-resonantly driving population out of the $\ket{s}$ state.
In the experiments and simulations above, we used linearly polarised 397 nm light, orthogonal to the quantisation axis, which drives $\sigma^{+}$ and $\sigma^{-}$ transitions.
The 397 nm intensity was 0.05 I$_0$, where I$_0$ is the saturation intensity of the 397 nm transition.
Since the laser is centred on a $\sigma^{+}$ transition, if the $\sigma^{-}$ and/or $\pi$ polarisation component is minimised, the desired optical pumping process out of the $F=3$ state is reduced, whilst increasing the probability of $\sigma^{-}$ and/or $\pi$ photons exciting a $\ket{s}\rightarrow$4P$_{1/2}$ transition.
To illustrate this, we rotate the beam's linear polarisation axis to have a $45^\circ$ angle with respect to the quantisation axis (rather than $90^\circ$ as above).
This allows the 397 nm laser to off-resonantly drive $\pi$ transitions in addition to $\sigma^{+}$ and $\sigma^{-}$ transitions.
In this case we measure an error of $1.9(3) \times 10^{-3}$ in preparing $\ket{s}$, which is close to our simulated error of $1.5 \times 10^{-3}$ and about 3 times worse than with $\sigma^+/\sigma^-$ polarisation.
Hence, although the FSSP scheme is still somewhat sensitive to polarisation, it is orders of magnitude less sensitive than the PSSP scheme.
In the implementation of FSSP described in Figs.~\ref{fig:main_fig2},~\ref{fig:main_fig3} and~\ref{fig:main_fig4}, the scheme is significantly slower (3 ms) than the PSSP (600 $\upmu$s), and is limited in speed by the MW operations.
Since the smallest detuning between a driven MW transition, and a MW transition involving the stretch state, is 240 MHz, the total MW pulsing duration could be lowered by approximately one order of magnitude, from 5 $\upmu$s to $\approx$ 500 ns per cycle (allowing for a $1 \times 10^{-4}$ off-resonant excitation error), giving a state preparation time of 300$\upmu$s.
Further speed-up could then be obtained through pulse shaping.
In practice, the on-chip resonator prohibits the driving of all MW transitions at sufficient amplitude, and we restrict the maximum injected MW power to 1 W per frequency tone, which ultimately limits the speed of our state-preparation scheme.

\begin{figure}
    \centering
    \includegraphics[width=0.45\textwidth]{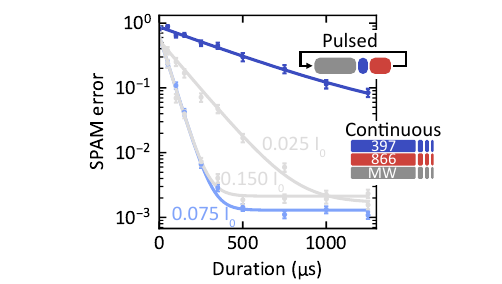}
    \caption{
        \textbf{Continuous-wave implementation of FSSP.}
        State-preparation and measurement (SPAM) error measured in a continuous implementation of FSSP (light blue, grey) and compared to the pulsed implementation (dark blue, described in Fig.~\ref{fig:main_fig2}).
        %
        % For each datapoint, data was acquired until the uncertainty in its value fell below $15\%$ of its nominal value.
        %
        The multiple datasets for the continuous case reveal the influence of 397 nm laser intensity on the state-preparation error and duration.
        At low intensity (grey, 0.025 I$_0$), convergence of the error is slower, with little advantage in error.
        When higher intensities are used (grey, 0.150 I$_0$), convergence occurs on a similar timescale, but the error is larger due to off-resonant excitation of the target state.
        We therefore empirically find an optimal intensity of 0.075 I$_0$ (light blue).
    }
    \label{fig:main_fig5}
\end{figure}

Since most of the microwave transitions are only driven for a fifth of the total cycle time, we also implement a ``continuous'' version of FSSP, where all laser and microwave fields are kept on in an attempt to speed up the state-preparation time.
However, in doing this, we can no longer benefit from comparing experimental data to the rate equation model.
As already identified in simulations of the pulsed implementation (see Sec.~\ref{sec:Simulation}), the intensity of the 397 nm light has a large influence on the process: a high intensity increases the error, and a low intensity slows down the state-preparation with marginal reduction in the error.
In the absence of a reliable numerical model, we sweep the laser intensity to experimentally determine the optimal setting.
The SPAM error for varying 397 nm light intensity is depicted in Fig.~\ref{fig:main_fig5}, highlighting the good balance between speed and error obtained with $I=0.075 I_0$.
The continuous implementation reduces the time required to prepare $\ket{s}$ from 3 ms to 750 $\upmu$s, at the expense of a small increase in $\ket{s}$ state-preparation error.
When combined with the transfer-pulses to the qubit state, the \textit{qubit} state-preparation and measurement error is measured to be $1.2(1) \times 10^{-3}$, and after substraction of the measured readout and transfer pulse errors ($4.7(4) \times 10^{-4}$), we attribute $7.5(13) \times 10^{-4}$ to state preparation of $\ket{s}$ through continuous FSSP.

In conclusion, we have introduced a frequency-selective state-preparation (FSSP) method that does not rely on pure laser polarisation and is applicable to various ion species with nuclear spin $I>1/2$ operated at ``intermediate-field''.
We report a SPAM error of $1.2 (1) \times 10^{-3}$ in $840\ \upmu$s using this method.
We compare this to our implementation of a polarisation-selective state-preparation method, with which we obtain a SPAM error of $4.7(4) \times 10^{-4}$ in $1.2$ ms.
The latter is an improvement on the previous state-of-the-art~\cite{harty2014} for intermediate-field $^\text{43}$Ca$^\text{+}$ ($6.8(5) \times 10^{-4}$ SPAM error).
In our implementation of the FSSP, both error and state-preparation duration are limited by the design of our surface trap chip which features an on-chip resonator.
Without resonant suppression of certain transitions, we would be able to raise the microwave driving rate, enabling us to lower the pulse duration by at least an order of magnitude without impacting the error, corresponding to $<300\ \upmu$s and $<75\ \upmu$s for the pulsed and continuous implementations respectively.
We would also be able to transfer the $\ket{F=4,M=-4}$ state to the qubit without being overly affected by decoherence in the transfer pulses, and so, by preparing this state with FSSP rather than $\ket{F=4,M=4}$, could theoretically lower the error to $\sim 1\times10^{-4}$.
Moreover, fidelity could be enhanced by transitioning to heavier ions, thereby increasing the hyperfine splitting and reducing the fundamental error source: off-resonant driving bridging the hyperfine splitting.
This scheme will simplify the scaling of ion trap based quantum computers using integrated optics, where purely circularly-polarised light is difficult to engineer.
It also allows for simpler laser delivery systems to control ions, with fewer beam paths and free space optical components, and lesser requirements on the quantisation field direction.
Furthermore, this scheme could be applied to the state preparation of optical and metastable state qubits~\cite{Allcock2021}.

\begin{table}
    \centering
        \begin{tabular}{|c|c|c|c|c|c|}
         \hline
         Pol.&Method&State  &Theory&Experiment&Duration\\
         $\sigma^+,\sigma^-,\pi$& &$\ket{F,M}$&[$\times 10^{-4}$]&[$\times 10^{-4}$]&[ms]\\
         \hline
         \hline
         $\sigma^+$& C \& P & $\ket{4,+4}$ & $- $ & $\lesssim0.1$ &0.6\\
         \hline
         $\sigma^+$/$\sigma^-$&P& $\ket{4,+4}$ & 6.3 & $6.1(11)$ &3\\
         \hline
         $\sigma^+$/$\sigma^-$&C& $\ket{4,+4}$ & - &  $7.5(13)$& 0.75\\
         \hline
         $\sigma^+$/$\sigma^-$&P& $\ket{4,-4}$ & 1.3 &-&-\\
         \hline
         $\sigma^+$/$\sigma^-$/$\pi$& P & $\ket{4,+4}$ & $15$ & $19(3)$&15\\
         \hline
        \end{tabular}
        \\
\caption{
    % \textbf{Summary of state preparation error.}
%
This table summarises the state preparation errors presented in this Letter (excluding transfer pulses, see Table~\ref{table:error_budget}).
In the method column, ``C'' and ``P'' correspond to continuous and pulsed schemes respectively.
The durations correspond to the experimental implementations.
}
\label{table:error_summary}
\end{table}

\vspace{5mm}
\textbf{Acknowledgments:}
This work was supported by the U.S. Army Research Office (ref. W911NF-18-1-0340) and the U.K. EPSRC Quantum Computing and Simulation Hub (ref. EP/T001062/1).
M.F.G. acknowledges support from the Netherlands Organization for Scientific Research (NWO) through a Rubicon Grant.
A.D.L. acknowledges support from Oxford Ionics Ltd.

%% file: SI.tex
%!TEX root = arxiv.tex

\begin{figure*}
    \centering
    \includegraphics[width=0.9\textwidth]{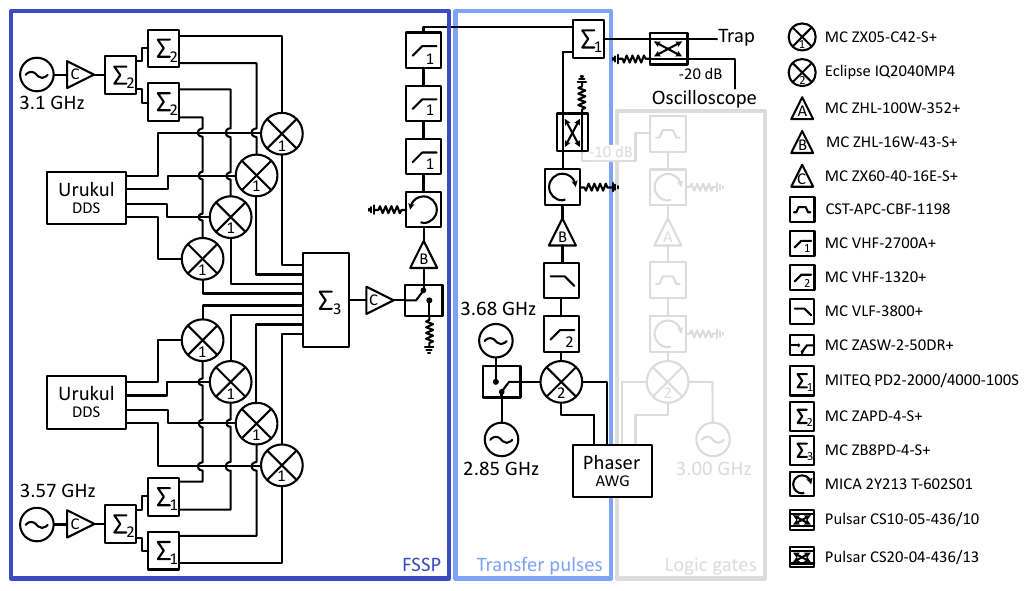}
    \caption{
    \textbf{Microwave setup.}
    On the left we show the circuitry used to drive hyperfine transitions during FSSP, in the center, the setup used to drive the transfer pulses used in readout and both approaches to state preparation, and the shaded portion on the right remains unused in the experiments presented here, but is otherwise used for driving logical operations.
    }
    \label{fig:microwave_setup}
\end{figure*}

\section{Microwave setup}
\label{sec:microwave_setup}

The microwave (MW) setup is depicted schematically in Fig.~\ref{fig:microwave_setup}.
During FSSP, MW pulses are used to transfer all states within the S$_{1/2}$ manifold (except $\ket{s}$) to $\ket{F=3,M=+3}$.
Ideally $\pi$-pulses would be used, but by instead employing short and strong pulses (with Rabi-frequencies $\gtrapprox$MHz), we are able to simultaneously drive transitions $M \rightarrow M+1$ $M +1\rightarrow M$ (split by $\approx$160 kHz), thereby reducing the total number of frequency tones required from 14 to 8.
We utilise two 4 channel DDS-based frequency synthesisers ``Urukuls'' from the Sinara~\cite{Sinara} hardware ecosystem to generate square pulses up to 400 MHz, which are upconverted and then combined.
The local oscillator frequencies used in the upconversion are chosen such that one of the sidebands of the mixing process is far detuned from any hyperfine transition, particularly transitions which would affect the $\ket{s}$ state.
After pre-amplification, we employ a switch to suppress leakage when this part of the microwave chain is not in use.
After amplification, high-pass filters are used to suppress microwaves at the ($\ket{s} \leftrightarrow \ket{F=3,M = 3}$) transition frequency, which would affect the FSSP.
To transfer the $\ket{s}$ state to and from the qubit, we use an arbitrary waveform generator (AWG) ``Phaser'' from the Sinara hardware ecosystem~\cite{Sinara}, to generate the I and Q quadratures of $\approx1\ \upmu$s $\text{sin}^2$ shaped pulses at up to 400 MHz.
These pulses are up-converted using an IQ-mixer and a choice of two local oscillators, enabling access to the full range of hyperfine transitions within the S$_{1/2}$ manifold, followed by filtering and amplification.
The microwave properties of the surface trap driven by this setup are detailed in Ref.~\cite{weber2022cryogenic}, in particular the characteristics of the on-chip resonator which plays an important role in the performance of the FSSP characterisation.

\begin{figure*}
    \centering
    \includegraphics[width=0.9\textwidth]{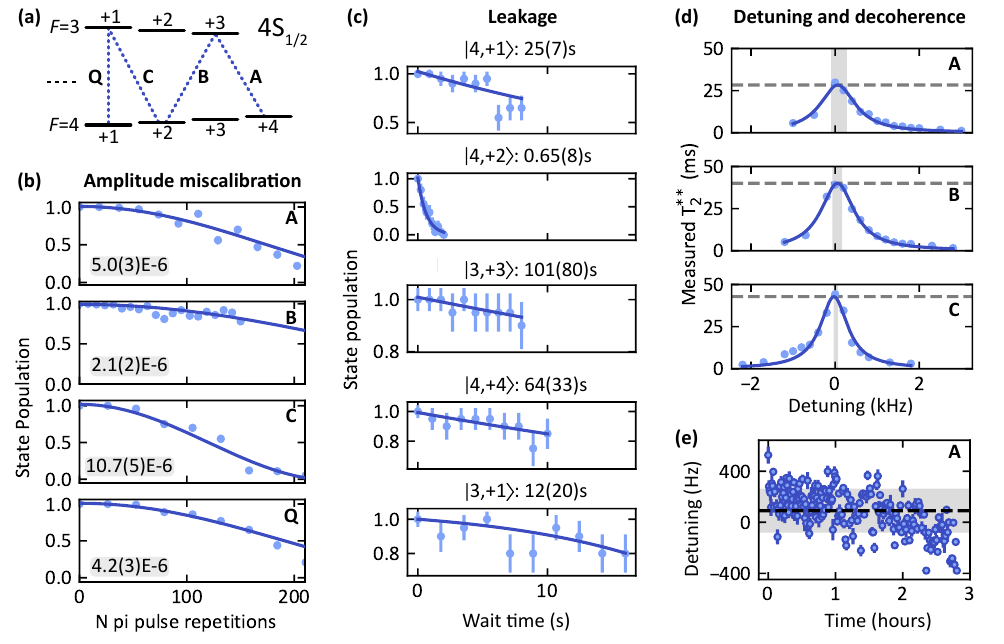}
    \caption{
        \textbf{Transfer pulse error characterisation.}
        \textbf{(a)} Ground state manifold of the $^{43}$Ca$^{+}$ ion at high field (28.8 mT), and the MW transitions used in the transfer to/from the qubit states (dotted lines).
        \textbf{(b)} \textbf{Amplitude miscalibration measurement.}
        For each transition, we perform a varying even number of $\pi$-pulses, expecting the population to return to the initial state in the case of ideal pulses.
        If there is a miscalibration of the pulse amplitude, we observe a Rabi oscillation with increasing number of pulses $N$.
        Fitting a sinusoid (dark blue) through the data lets us determine the error of a single $\pi$-pulse, which is annotated in the grey box in each plot.
        \textbf{(c)} \textbf{Leakage characterisation.}
        We prepare one of the states occupied during the transfer pulses and then wait for a varied delay before reading out the initial state occupation (data in light blue).
        The decay in initial population is fitted with an exponential (dark blue).
        This provides us with an estimated lifetime for each state, annotated on the top of the plots.
        \textbf{(d)} \textbf{Detuning and decoherence measurements.}
        For each transition, we perform randomised benchmarking (RB) for a range of inter-pulse delays.
        At the short timescales, the RB error scales approximately linearly with the delay, from which we extract a short-time-scale coherence time, or $T_2^{**}$.
        A frequency detuning between the MWs and the driven transition will lower the effective $T_2^{**}$, and so we repeat the measurement for different detunings (light blue).
        A Lorentzian is fitted through the resulting data (dark blue), and the maximum $T_2^{**}$ is indicated by the grey dashed line.
        This can be taken as an accurate measure of $T_2^{**}$, given the typical deviation in detuning, indicated with the shaded grey region, determined by the measurement of panel (e).
        \textbf{(e)} \textbf{Detuning drift.}
        The frequency of transition A was measured repeatedly over three hours.
        The black dashed line shows the mean detuning and the light grey area shows the standard deviation.
        Each measurement consisted of a detuning scan in a Rabi-oscillation experiment, fitted with a Lorentzian to extract the frequency.
    }
    \label{fig:transfer_pulse_error}
\end{figure*}

\section{Transfer pulse errors}
\label{sec:transfer_pulses}

\textbf{Leakage:}
Implementing transfer pulses with different frequencies using a single ``Phaser'' AWG requires updating the internal numerical oscillator between pulses, which can take up to 150 $\upmu$s.
On this time-scale, the undesired population transfer away from a given state, attributed to microwave and laser power leakage and therefore referred to as ``leakage'' error, becomes significant.
This is measured by preparing a given state, idling for a varying duration, then reading out the state occupation.
Fig.~\ref{fig:transfer_pulse_error}(c) shows a measurement of the resulting ``lifetimes'' for the different states occupied during the transfer pulses.
We use this to optimise the transfer pulse process by minimising the idle time in the worst affected states, and to compute an estimate of the total leakage error, provided in Table~\ref{table:error_budget}.

\textbf{Amplitude miscalibration:}
To quantify a transfer pulse miscalibration, the ion is subjected to 2$N$ pulses which, in the ideal case, would leave the state unaffected.
A miscalibration of the pulse area will result in Rabi-oscillations with increasing $N$, and the ``period'' (given in number of pulses) informs us about the error.
These diagnostic measurements are shown in Fig.~\ref{fig:transfer_pulse_error}(b), from which we infer the amplitude miscalibration error reported in Table~\ref{table:error_budget}.

\textbf{Detuning and decoherence:}
In this section we discuss the characterisation of two errors arising from uncertainty in the frequency of the transfer pulse transitions.
Firstly decoherence, attributed to fast magnetic field noise, and secondly frequency offsets, caused by a miscalibration or slow drift in the magnetic field strength.
Since the transfer pulse duration ($\sim1\upmu$s) is much shorter than the decoherence time ($\sim$ 10ms), the most accurate measure of its impact will be given by memory benchmarking \cite{Sepiol2019}.
This involves performing randomised benchmarking (RB)~\cite{Knill2008RBM} measurements with varying inter-pulse delays to infer a timescale for decoherence, referred to as $T_2^{**}$~\cite{leu2023}.
An offset or drift in the transition frequency will effectively lower the measured $T_2^{**}$ and so, to separate detuning errors from decoherence, we perform memory benchmarking for varying detunings as shown in Fig.~\ref{fig:transfer_pulse_error}(d).
To characterise the error from the frequency offsets, we measure the frequency of transition A (the most sensitive to magnetic field changes) over three hours, as shown in Fig.~\ref{fig:transfer_pulse_error}(e) and conclude that slow frequency drifts will have a negligible impact on the total SPAM error.

\section{Optical readout errors}
\label{sec:optical_readout}

In this section, we discuss the errors stemming from the optical readout process.
After mapping the qubit states $\ket{F=3,M=+1}$ and $\ket{F=4,M=+1}$ to $\ket{F=4,M=+4}$ and $\ket{F=3,M=+1}$ respectively, we implement a ``shelving'' procedure, followed by a fluorescence measurement.
For shelving, we use 10 cycles of a laser pulse sequence consisting of a $\sigma ^{+}$-polarised 393 nm pulse, two $\sigma^{+}$-polarised 850 nm pulses and a $\pi$-polarised 850 nm pulse to transfer $\ket{F=4,M=+4}$ to the 3D$_{5/2}$ manifold~\cite{myerson2008}.
Finally, we use 397 nm and 866 nm light to measure the fluorescence of the population in the 4S$_{1/2}$ manifold.
% Aaron is this the theory value or the
We estimate the total error of the optical readout to be $1.4(1) \times 10^{-4}$ for the dark state and $2.8 (1)\times 10^{-4}$ for the bright state, with errors originating from the following sources.

\textbf{Off-resonant shelving:}
One error source stems from the shelving of the ``bright'' state $\ket{F=3,M=+1}$ by the detuned 393 nm laser pulses.
We estimate the magnitude of this error theoretically using our experimental parameters: a total of 60 $\upmu$s exposure to 393 nm laser light at an intensity of $0.03$ saturation intensities, resonant with the $\text{S}_{1/2}^{4,+4} \leftrightarrow \text{P}_{3/2}^{5,+5}$ transition.
The resulting error of $2.5 \times 10^{-4}$ dominates the optical readout error of the ``bright'' state.

\textbf{Thresholding:}
The inherently statistical nature of photon emission can lead to an insufficient number of photon detection events to correctly identify the ion in the bright state.
To quantify this error, we measure a bright state repeatedly, and find an error of 3.4(10) $\times 10^{-5}$ in successfully identifying the ion as bright.
Decreasing the threshold number of photons would reduce this error, at the expense of increasing the error from detecting photons which were not scattered by the ion.
Indeed, 397nm light which is not scattered by the ion, most notably scattered by the surface trap chip, can reach our photon detector.
This light can be detected as a bright state of the ion, but with a small probability.
The threshold number of photons for a bright state is chosen to minimise the sum of these two error sources, such that the error from measuring such scattered photons is also $3.4 \times 10^{-5}$.
\textbf{Shelving failure:}
In the case of dark state detection, the probability of shelving approaches 1 exponentially with the total 393 nm pulse duration, which decreases the readout error.
In the meantime however, deshelving can occur spontaneously, and to an unknown state in S$_{1/2}$, due to the $\approx 1$ s lifetime 3D$_{5/2}$ states.
These two effects are balanced to minimise the probability of failing to shelve, producing an error of $4 \times 10^{-5}$.
%

%Formula: 1/2*t_fluo/T_lifetime
% this formula comes from integrating the prob of decaying in dt (P=dt/T_lifetime) times the error (err=(1-t/t_fluo),assumes linear error)
\textbf{Deshelving:}
Similarly, there is a finite probability of deshelving during the fluorescence readout, resulting in photons being scattered on the ion when the ion would ideally remain dark.
The resulting error is $7.6 \times 10^{-5}$.

\section{FSSP simulation}
\label{sec:Simulation}
\begin{figure}
    \includegraphics[width=0.45\textwidth]{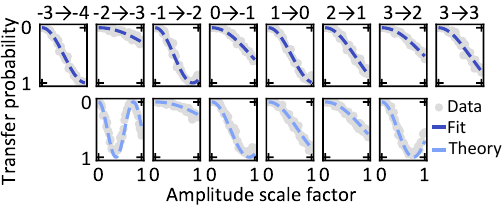}
    \caption{
    \textbf{FSSP MW characterisation.}
    Each panel shows the probability of transferring one state to another, as the microwave amplitude is scaled from 0 to the amplitude used in the experiment.
    Each column is annotated with two $M$ numbers as follows: $p\rightarrow q$.
    The upper row corresponds to measurements where the $F=3,M=p$ prepared, and transfer to $F=4,M=q$ is measured.
    Conversely, the bottom row corresponds to measurements where the $F=4,M=p$ prepared, and transfer to $F=3,M=q$ is measured (driven by the same microwave tone as the measurement above).
    In the upper row, the oscillation is fitted to determine the relationship between microwave pulse amplitude and Rabi frequency.
    In the bottom row, we show the expected Rabi-oscillation based on the extracted Rabi frequency, the detuning between the upper and lower transition, and the theoretical difference in magnetic dipole matrix elements.
    }
    \label{fig:transition_turns}
\end{figure}

\begin{figure}
    \includegraphics[width=0.45\textwidth]{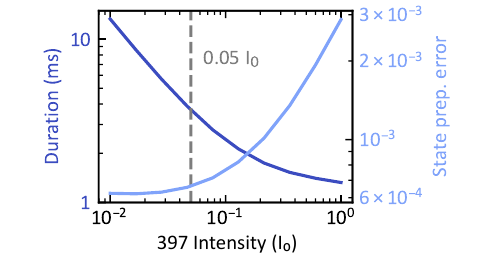}
    \caption{
    \textbf{397 nm pulse duration/infidelity compromise.}
    We simulate the FSSP pulse scheme for an increasing numbers of cycles until an error reduction per cycle of less than $1 \times 10^{-7}$ is reached.
    The resulting state preparation error (dark blue) and the duration (light blue) is shown as a function of the 397 nm intensity.
    The grey dashed line shows the intensity used in our experiments.
    }
    \label{fig:duration_v_error}
\end{figure}

In this section we discuss the simulations of the FSSP in Figs.~\ref{fig:main_fig2},\ref{fig:main_fig3}.
We make use of an open-source package, ``atomic-physics''~\cite{atomic_physics}, to compute the rate equation governing state occupations when the ion is driven by a single laser beam.
The rate equation describes the evolution of the (classical) probabilities to occupy given states, i.e. a vector of $N$ real numbers, where $N$ is the number of atomic states involved.
This approach is considerably more tractable than solving a more general master equation, such as the Lindblad equation, which could be necessary if three or more states were coupled with different driving terms, giving rise to an $N^2$ sized density matrix and possibly shorter timescale dynamics.
The microwave operations are characterised experimentally, and converted into a $N^2$ matrix acting on the vector of occupation probabilities.
This model is used to predict state-preparation errors and durations and determine optimal pulse schemes, including optimal laser pulse durations and intensities.
Our characterisation of the microwave pulses consists of measuring Rabi-oscillations on the different transitions with varying pulse amplitudes, as shown in Fig.~\ref{fig:transition_turns}.
In these measurements, we drive Rabi-oscillations using the same multi-frequency pulse as in the FSSP scheme to capture the power-induced non-linearities of the drive chain.
For pulses that simultaneously drive the two near-degenerate transitions $M \rightarrow M+1$ and $M +1\rightarrow M$, the detuning and different matrix elements translate to two different Rabi-oscillations.
We adjust the microwave amplitudes used in order to avoid performing $\sim 2\pi$ pulse on any transition, even if this means reducing the transfer probability of its near-degenerate counterpart.
In simulation, we find that this combination of transfer probabilities only reduces the state-preparation fidelity by $\approx 1\times 10^{-4}$ with respect to the ideal case of applying $\pi$ pulses to all transitions.
This ideal situation can theoretically be achieved using the composite pulse scheme presented in Ref.~\cite{leu2023}, exploiting the fact that the degenerate transitions have different interaction strengths due to their different matrix elements.
Using this characterisation of the microwave pulses, an optimal 397 nm pulse intensity and duration can be determined in simulation (the 866 nm pulse settings have a much smaller impact on the error).
Fig.~\ref{fig:duration_v_error} shows the trade-off between state-preparation duration and error that comes from the choice of 397 nm laser intensity for the pulse duration (150 ns) we use.
Our choice of 0.05 saturation intensities used in the measurements shown in Fig.~\ref{fig:main_fig3} minimises the state-preparation error at the expense of a longer duration.


\begin{figure}
    \includegraphics[width=0.45\textwidth]{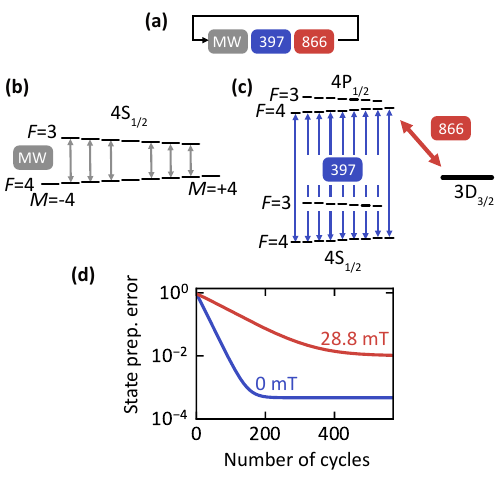}
    \caption{
    \textbf{Alternative scheme, at high and low magnetic fields.}
    This figure illustrates the pulse sequence used for the simulation of the state preparation scheme presented in Ref.~\cite{quantinuumStatePrep}.
    \textbf{(a)} Pulse sequence of MW (grey), 397 nm laser (blue), and 866 nm laser (red) pulses, used for the state preparation scheme of Ref.~\cite{quantinuumStatePrep}.
    \textbf{(b)} MW transitions ($\pi$-pulses) driven within the 4S$_{1/2}$ manifold.
    \textbf{(c)} Optical transitions driven by the multiple tones of 397 nm laser light.
    An 866 nm laser recovers population that decays to the 3D$_{3/2}$ manifold.
    \textbf{(d)} Simulated state preparation errors for the clock qubit at low- and high-field, motivating the use of the alternative scheme presented in this work for state preparation at high field.
    }
    \label{fig:state_prep_sim}
\end{figure}

\section{Comparison of high- and low-field}
\label{sec:comparison}
Here we show how the frequency-selective state-preparation method of Ref.~\cite{quantinuumStatePrep} leads to large errors when applied to high-field hyperfine qubits.
To do so, we use the theoretical approach of Sec.~\ref{sec:Simulation} to model the method of Ref.~\cite{quantinuumStatePrep}.
Our theoretical implementation of this scheme is summarised in Fig.~\ref{fig:state_prep_sim} (a-c).
The scheme also exploits the frequency selectivity of microwave fields and a 397 nm laser, with one major difference in that multiple 397 nm transition have to be addressed.
By optimising the laser pulse parameters, we are able to qualitatively reproduce the expected results for low-field $^{43}$Ca$^{+}$, namely a state-preparation error in the $10^{-4}$ regime, see Fig.~\ref{fig:state_prep_sim} (d).
However, at high fields (28.8 mT in our case), the energy splitting between the $\Delta F = 0,\Delta M = 1$ states is $\approx$100 MHz, which is larger than the linewidth of 397 nm transitions.
Whereas in the low-field case each 397 nm frequency component off-resonantly contributes to optically pumping every state in the $F=3$ manifold, effectively reducing the 397 nm intensity required in each transition, the high-field case requires an increased intensity, or alternatively an increased number of cycles for the same intensity.
The high-field case therefore requires a higher exposure to 397 nm light, increasing the impact of off-resonant driving out of the target state, and resulting in a larger error.
This is illustrated in Fig.~\ref{fig:state_prep_sim}(d), where we show the state-preparation result for an illustrative choice of laser pulse settings in at high-field.